\documentclass[aps,prx,preprint,superscriptaddress,amsmath,amssymb]{revtex4-2}

\usepackage{graphicx}
\usepackage{dcolumn}
\usepackage{bm}
\usepackage[colorlinks,linkcolor = blue,
citecolor = blue,urlcolor=blue]{hyperref}

\begin{document}

\title{Efficiently Laser Driven Terahertz Surface Plasmon Polaritons on Long Metal Wire}

\author{Shuo-ting Shao}
\thanks{These authors contributed equally: Shuo-ting Shao, Xiang-bing Wang.}
\affiliation{School of Nuclear Science and Technology, CAS Key Laboratory of Geospace Environment, University of Science and Technology of China, Hefei, China}
\author{Xiang-bing Wang}
\thanks{These authors contributed equally: Shuo-ting Shao, Xiang-bing Wang.}
\affiliation{School of Nuclear Science and Technology, CAS Key Laboratory of Geospace Environment, University of Science and Technology of China, Hefei, China}
\author{Rong Huang} 
\affiliation{Key Laboratory for Laser Plasmas (Ministry of Education), School of Physics and Astronomy, Shanghai Jiao Tong University, Shanghai 200240, China}
\author{Guang-yue Hu}
\thanks{\textcolor{blue}{Corresponding author: gyhu@ustc.edu.cn}}
\affiliation{School of Nuclear Science and Technology, CAS Key Laboratory of Geospace Environment, University of Science and Technology of China, Hefei, China}
\author{Min Chen}
\thanks{\textcolor{blue}{Corresponding author: minchen@sjtu.edu.cn}}
\affiliation{Key Laboratory for Laser Plasmas (Ministry of Education), School of Physics and Astronomy, Shanghai Jiao Tong University, Shanghai 200240, China}
\author{Hui-bo Tang}
\affiliation{School of Nuclear Science and Technology, CAS Key Laboratory of Geospace Environment, University of Science and Technology of China, Hefei, China}
\affiliation{School of Physics, Harbin Institute of Technology, Harbin 150001, China}
\author{Long-yu Kuang}
\affiliation{Research Center of Laser Fusion, China Academy of Engineering Physics, Mianyang, Sichuan 621900, China}
\author{Yu-xi Liu}
\affiliation{School of Nuclear Science and Technology, CAS Key Laboratory of Geospace Environment, University of Science and Technology of China, Hefei, China}
\author{Yu-qiu Gu}
\thanks{\textcolor{blue}{Corresponding author: yqgu@caep.ac.cn}}
\affiliation{Research Center of Laser Fusion, China Academy of Engineering Physics, Mianyang, Sichuan 621900, China}
\author{Yong-kun Ding}
\thanks{\textcolor{blue}{Corresponding author: ding-yk@vip.sina.com}}
\affiliation{Research Center of Laser Fusion, China Academy of Engineering Physics, Mianyang, Sichuan 621900, China}
\author{Hong-bin Zhuo}
\affiliation{Shenzhen Key Laboratory of Ultraintense Laser and Advanced Material Technology, Center for Advanced Material Diagnostic Technology, and College of Engineering Physics, Shenzhen Technology University,
Shenzhen 518118, China}
\author{Ming-yang Yu}
\affiliation{Shenzhen Key Laboratory of Ultraintense Laser and Advanced Material Technology, Center for Advanced Material Diagnostic Technology, and College of Engineering Physics, Shenzhen Technology University,
Shenzhen 518118, China}

\date{\today}

\begin{abstract}
We experimentally demonstrate a novel scheme for efficiently generating intense terahertz (THz) surface plasmon polaritons (SPPs) on a sub-wavelength-diameter meter-long metal wire. Driven by a \emph{subrelativistic} femtosecond laser ($a_0=0.3$, 3 mJ) focused at the wire's midpoint, single-cycle ten-megawatt THz SPPs are excited and propagating bidirectionally along it over 25 cm. The measured laser-to-SPPs energy conversion efficiency is reaching up to $\sim2.4\%$, which is the highest value at present. It is proved that the THz SPPs are excited by coherent transition radiation of the subrelativistic laser produced escaping electrons. Particle-in-cell together with CST simulations confirm the experimental observations. Our scheme of using readily available subrelativistic laser should thus be useful to applications requiring terawatt level single-cycle THz SPPs.
\end{abstract}


\maketitle

\section{Introduction}
Intense THz sources\cite{1.np2017,2.jpd2017,3.j02016} can be essential for exploring condensed-matter physics\cite{4.n2012,5.np2013,6.prl2015}, biosensing\cite{7.pmb2002,8.pb2012,9.sr2016,10.jbo2003}, wireless communication\cite{11.NaturePhoton2013}, etc. Traditionally, generation of intense free-space THz radiation relies on accelerator-based relativistic electron beams that require very large facilities\cite{12.jopb2013,13.prl1991,14.pre1995,15.Nature2002,16.NaturePhys2008,17.prl2009,18.prl2011,19.np2019}. Although tabletop lasers can overcome some of the limitations, THz sources based on laser optical rectification\cite{20.oe2006,21.ol2008,22.apl2011,23.NC2015} and photoconductive antennas\cite{24.apl1984,25.nc2013} suffer from crystal damage. Since plasma is not damaged by intense light within the timescale of most applications, it has been invoked for generating intense THz pulses\cite{26.IEEE2019}. For example, broadband THz pulses can be generated by laser filamentation in gases\cite{27.prl2006,28.prl2007,29.np2008,30.prl2013} or liquids\cite{31.nc2017,32.apl2017}. Other intense laser-plasma interaction based schemes using gas\cite{33.prl2003,34.pre2004} or solid\cite{35.Physrl1993,36.111prl2013,37.114psrl2015,38.14njp2012,39.116prl2016,40.94pre2016,41.20njp2018,42.98pre2018,43.203pre2019,44.204pre2019} targets have also been proposed, and terawatt THz pulses with multi-millijoule energies have been demonstrated\cite{45.pnas2019,46.prx2020}. However, these intense free-space THz pulses suffer from diffraction loss and are accessible only in the line of sight, thus require large optical components for focusing and guiding.

Surface plasmon polaritons (SPPs) based on metal wire THz waveguide, which are coherent subwavelength electron oscillations localized near metal wire interfaces\cite{01Nature,02Zhang2020TerahertzSP,03,04.bookSPN,05}, allow accumulation and channelling of THz energy, and have strong potential for realizing compact subwavelength THz photonics\cite{47.nature2004,48.JOSAB2005,50.NC2016,51.PRB2008, 52.Sci.Rep.2015,49.PRE2018}. However, efficient coupling of metal wire-based SPPs with free-space THz radiation remains difficult\cite{49.PRE2018}. Recently, mechanisms based on laser driven hot electrons have been proposed for directly exciting THz SPPs or free-space radiation on metal wires\cite{ 52.Sci.Rep.2015}, including wire-guided helical undulators\cite{53.11NP2017,66.28OE2020}, SPPs amplification\cite{55.nature2022}, and current-carrying line antennas based on guided electrons \cite{54.PRE2017} or return currents\cite{JianshuoWang2024RadiationDA}. Notably, all mechanisms indicate a preference for the use of relativistic lasers to obtain intense THz. Moreover, there is still a lack of efficient generation and guiding mechanisms for practical applications. 

In this paper, we demonstrate a novel scheme to efficiently generate single-cycle intense THz SPPs by focusing a \emph{subrelativistic} femtosecond laser on the midsection of a meter-long subwavelength-diameter metal wire. For the first time, bidirectional electron and intense THz SPPs propagation along the wire, and emission of single-cycle THz radiation at the wire ends are observed, and we showed that subrelativistic, rather than relativistic (as invoked in all existing works), lasers are more suitable for exciting THz SPPs efficiently. This diametrically different result is attributed to coherent transition radiation (CTR) of the subrelativistic laser driven subrelativistic electrons from the long thin wire, and is confirmed by particle-in-cell and CST simulations.

\section{Experimental Setup}

\begin{figure}[ht]
\centering
\includegraphics[width = 0.9\textwidth]{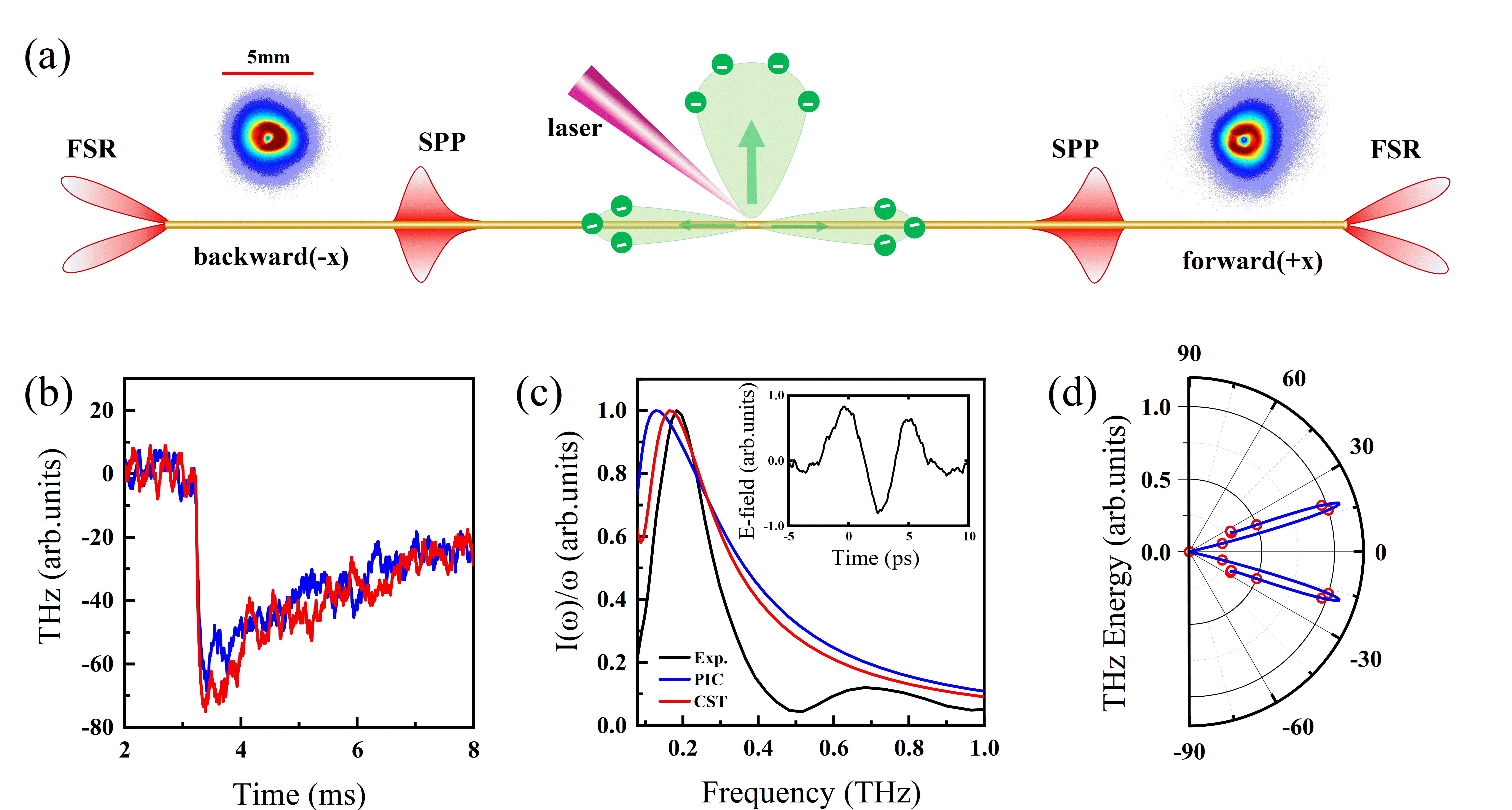}
\caption{\label{fig:1}(a) Result-based schematic of the experiment setup. A 3mJ, 40fs, subrelativistic ($a_0=0.3$) $p$-polarized femtosecond laser irradiates the top-center of a long metal wire to expel hot electrons, generate forward and backward (in the $\pm x$ directions, respectively) propagating THz SPPs and electron bunches (insets, using rainbow colormap), as well as free-space radiation (FSR) from the wire ends. The corresponding local electric field intensities are shown in shaded red color. (b) Energy signals of the THz radiation at the wire ends as detected using pyroelectric detectors. (c) Spectra and temporal profile (inset) of the THz SPPs from the experiment, PIC simulation, and CST simulation. (d) Angular distribution of the THz emission in the horizontal plane at the wire ends as detected.}
\end{figure}

The experiments are performed in a vacuum ($10^{-3}$ Pa pressure) chamber using a Ti:sapphire chirped-pulse amplified laser at the University of Science and Technology of China. As illustrated in Fig.~\ref{fig:1}(a), a 3 mJ, 40 fs, 790 nm, \emph{p}-polarized subrelativistic laser pulse incidents the top center of a metal wire with 10 $\mu$m (FWHM) focal spot and $10^{17}$ $W/cm^2$ intensity (or normalized field intensity $a_0=0.3$).
The laser contrast ratio (CR) in the nanosecond range is $10^{-8}$, and each femtosecond main pulse is preceded by a $\sim40$ fs prepulse with CR $\sim10^{-6}$ at 7 ns and a 100-500 ps-scale amplified spontaneous emission (ASE) pulse with CR $\sim4\times10^{-3}$ at 3 ns before the main pulse. The default laser incidence angle is $0^\circ$, or normal to the wire. Unless otherwise stated, tungsten wires of 30 $\mu$m diameter and 50 cm length are used. For comparison, experiments using laser incidence angles $30^\circ$, $45^\circ$, and $60^\circ$, wire diameters 50-500 $\mu$m, wire length 1 m, as well as wire material copper, have also been carried out.

The temporal profiles of the THz electric field and energy of the free-space THz radiation at the two ends of the wire are obtained using single-shot electro-optics pump-probe technique with 1 mm thick ZnTe crystals and 15 ps chirped probe laser\cite{56.611nature2022}, as well as pyroelectric detectors (Gentec-EO: THZ5B-BL-DA). The polarization and angular distribution of the THz pulses are measured using THz polarizer and continuously-variable aperture combined with pyroelectric detector. High-resistance silicon wafers are placed in front of the detectors to block the infrared scattering of laser. A TPX (Polymethyl Pentene) lens pair is used to collect the free-space THz radiation emitted from the wire ends and focus them onto the detectors.

The spatial distribution and energy spectra of the hot electrons at 12cm from the laser incidence spot (LIS) guided along the wire are measured using $22\times$22 mm$^2$ fluorescent screens (DRZ-Std) and electron magnetic spectrometer, respectively. The initial angular distribution of the laser expelled hot electrons is obtained at $\sim$5 $cm$ from the LIS using a large imaging plate (Fujifilm SR)\cite{57.84rsi2013}, covering $-110^\circ$ to $+110^\circ$ from the wire normal. The imaging plate is enclosed by 10 $\mu$m thick Al foil to block low-energy x rays and scattered light.

\section{Experimental Results}

\begin{figure}[ht]
\centering
\includegraphics[width = 0.9\textwidth]{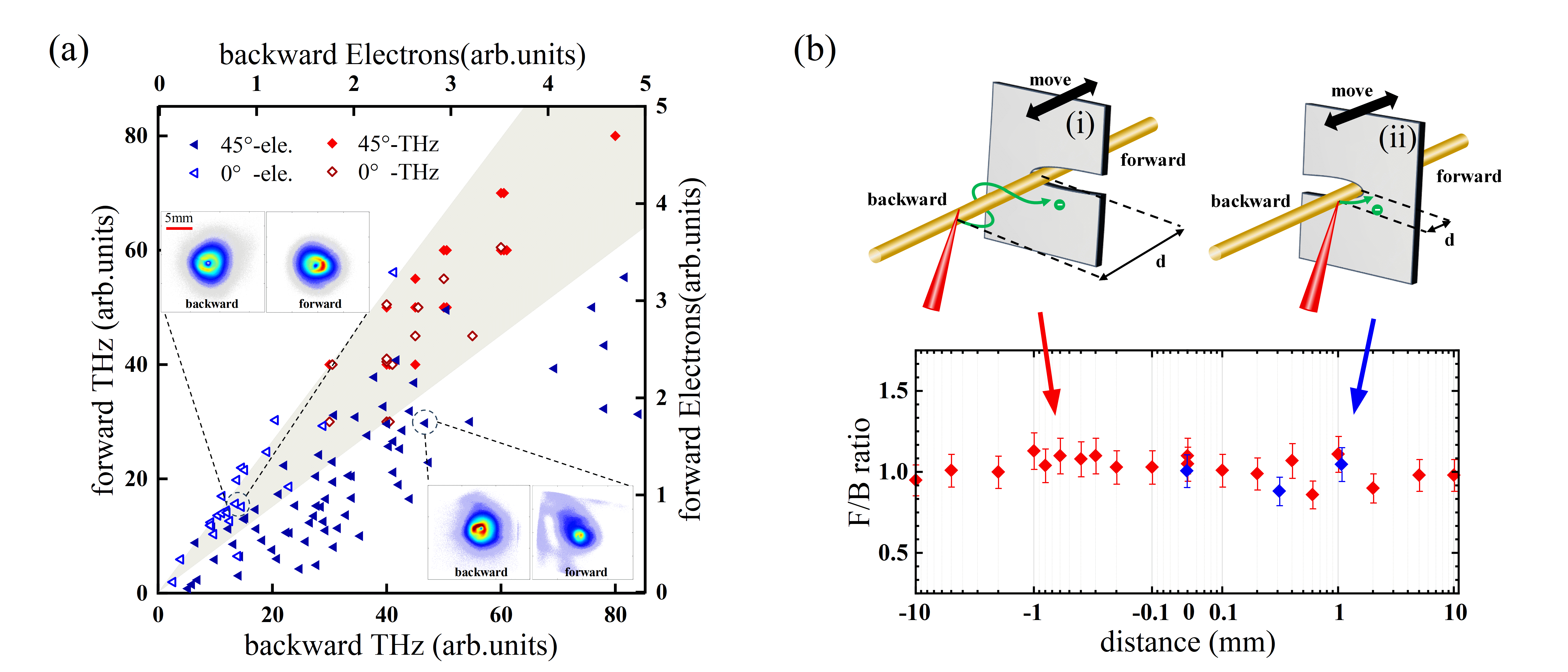}
\caption{\label{fig:2}(a) Energy and charge of the forward and backward ($+x$ and $-x$ directions, respectively) propagating THz SPPs and wire guided electrons, respectively. The shaded area shows the range ($1\pm0.25$) of the forward-to-backward (F/B) ratio. For $0^\circ$ laser incidence, there are nearly the same numbers of the forward ($+x$) and backward ($-x$) propagating SPPs and electrons. For $+45^\circ$ laser incidence the same remains true for the SPPs, but not for the electrons. Typical distribution patterns (rainbow colorscale, arbitrary units) of the electrons for $0^\circ$ and $+45^\circ$ laser incidence are shown in the dashed-lines guided upper-left and lower-right insets, respectively. (b) Ratio of the energy of the F/B propagating SPPs versus the distance between the blocking sheet and the LIS. The panels (i) and (ii) are for electron blocking on radially the opposite side (red) and same side (blue), respectively, of the LIS on the thin wire.}
\end{figure}

\textbf{Characteristics of THz SPPs.} As shown in Fig.~\ref{fig:1}(b) for the signals from the pyroelectric detectors, with normal laser incidence, the energy of the free-space THz radiation emitted at the ends of the wire has a typical symmetric distribution. Fig.~\ref{fig:2}(a) shows that the ratio of the THz radiation energy of the forward ($+x$)-to-backward ($-x$) (F/B) is within $1\pm0.25$ from shot to shot. The inset of Fig.~\ref{fig:1}(c) shows a typical temporal THz electric field profile: single cycle of $\sim5$ ps duration covering a broad range of frequencies up to $\sim1$ THz, with central frequency at $\sim0.2$ THz, or $\sim1.5$ mm wavelength. Fig.~\ref{fig:1}(d) for the THz radiation emitted from the wire ends shows that it has a hollow-cone angular distribution with $\sim20^\circ$ opening angle. The polarization detector output shows that the radiation is radially polarized, suggesting that it can be a Sommerfeld surface mode on the wire\cite{48.JOSAB2005,52.Sci.Rep.2015,55.nature2022}.
The total energy of the free-space THz radiation emitted at the wire ends from the pyroelectric and single-shot electro-optics measurements is $\sim2\times12\mu$J after accounting for transmittance loss in the TPX lens and silicon wafer, and the corresponding laser-to-THz radiation energy conversion efficiency is $\sim2\times0.4\%$. Considering that most (about 2/3) of the SPPs' power is reflected at the wire ends due to impedance mismatch, the laser-to-SPP energy-conversion efficiency can be up to $\sim2\times1.2\%$, which is the highest value at present(see Fig.~\ref{fig:5}(a), Fig.S3 and Table S1 in Supplementary Material). That is, the power of the SPPs on the wire can be more than 10 MW \cite{52.Sci.Rep.2015}.

\begin{figure}[ht]
\centering
\includegraphics[width = 0.9\textwidth]{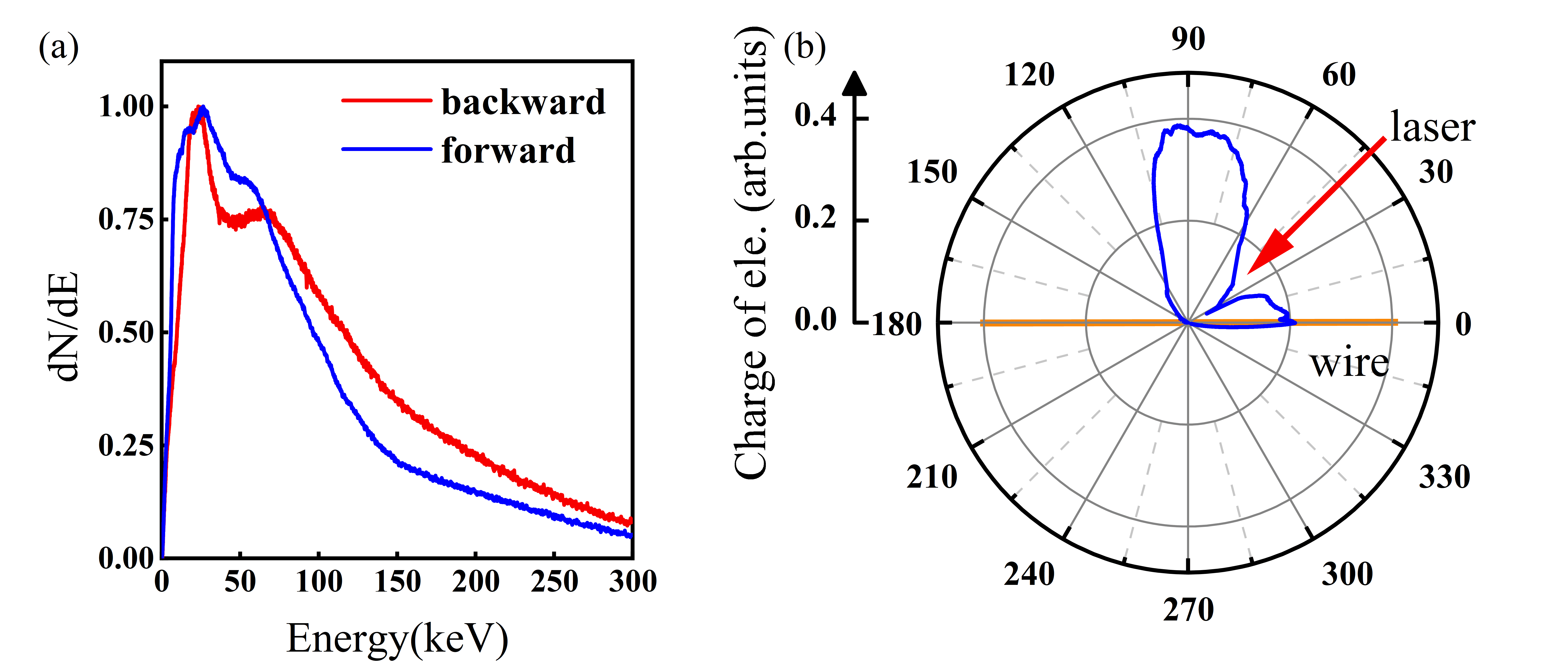}
\caption{\label{fig:3}Experiment results. (a) Electron energy spectra of the forward ($+x$) and backward ($-x$) guided hot electrons along the wire for normal ($0^\circ$) laser incidence. (b) Initial angular distribution of the hot electrons near the LIS for $45^\circ$ laser incidence, indicating most of the laser-expelled hot electrons there have been emitted perpendicular to the wire.}
\end{figure}

\textbf{Features of guided hot electrons.} The insets of Fig.~\ref{fig:1}(a) shows that at $0^\circ$ laser incidence, the guided hot electrons distributions in the $\pm x$ directions are symmetric. One can see that at 12 cm from the LIS, the electrons are still localized in a ring area of outer diameter $\sim$5 mm, with the wire at the center. Such almost divergence-free guiding of laser expelled hot electrons by thin wires has been noted earlier\cite{58.432nature2004,59.106prl2011,60.110prl2013}.
Fig.~\ref{fig:3}(a) shows that these electrons have a broad ($\sim20-150$ keV) energy distribution, as to be expected of laser expelled hot electrons.

\textbf{Non-relevance between THz
SPPs and guided-hot-electrons.} Fig.~\ref{fig:2}(a) shows that even when the laser incidence angle is $+45^\circ$, the bidirectional THz SPPs still have symmetric distribution, but the guided electrons do not: there are clearly much more electrons propagating in the backward ($-x$) direction. 
The guided electrons' backward-to-forward (B/F) charge ratio is about $1.8:1$ on the average and can reach $8:1$. The difference between the electron and SPP distributions confirms that the SPPs are not excited by the guided electrons, as found in several studies \cite{55.nature2022,53.11NP2017,54.PRE2017,66.28OE2020}.

\textbf{Manipulating the guided hot electrons.} To ascertain this conclusion, additional experiments, in which the wire-guided hot electrons are artificially manipulated, have been performed. We first block the forward ($+x$) electrons with a 0.5 mm thick Teflon sheet placed at different locations $d$ from, and radially on the opposite side of, the LIS, as shown in the panel (i) of Fig.~\ref{fig:2}(b), and obtain the energy of the SPPs and spatial distribution of the guided electrons. When the Teflon sheet is at the rear of the laser spot, for normal laser incidence the number of forward ($+x$) moving hot electrons becomes only $1/10$ that without the Teflon sheet , but there is no change in that of the bidirectional THz SPPs and backward ($-x$) moving hot electrons. Similar behavior occurs when the Teflon sheet is placed 0 to $-10$ mm from LIS (When the Teflon sheet is placed exactly at x=0, the electrons moving in both directions disappear completely, while the bidirectional THz SPPs remain unchanged.). Thus, the wire-guided helical undulator model\cite{53.11NP2017,66.28OE2020} and mechanisms based on laser driven hot electrons on the target backside\cite{10.1063/5.0013415,PhysRevE.108.065211, 46.prx2020} for THz radiation generation are not applicable to our results.

For further verification, the Teflon sheet is placed at 20 $\mu$m -1 mm from (i.e., very close to, including sub-THz-wavelength distance), and radially on the same side of, the laser spot. The laser incidence angle is changed to $45^\circ$ to avoid the Teflon sheet interfering the laser beam. The panel (ii) of Fig.~\ref{fig:2}(b) shows that the forward ($+x$) hot electrons completely disappear, but the distribution patterns of the backward ($-x$) electrons and emitted THz pulses do not change. Thus, the models of SPPs amplification\cite{55.nature2022} and current-carrying line antenna based on guided hot electrons\cite{54.PRE2017} also do not apply to our results.

\textbf{Excluding current-carrying antenna models.} We also examined the applicability of the current-carrying antenna model based on return current\cite{JianshuoWang2024RadiationDA}. As laser irradiates and expels hot electrons from the wire, a transient charge separation field is built up and drives return currents to its ends. Radiation from the return currents can be described by the dipole-antenna model of Smith\cite{Smith1997AnIT}. The initial profile of the transient current is assumed to be $I_{s}(t)=I_0e^{(-t^2/2\tau_0^2)}$, where $I_0$ and $\tau_0$ are constants. The radiated dipole electric field is\cite{Smith1997AnIT}
\begin{align}
    E(r,t)&=\frac{\mu_0c}{4\pi r}(\frac{\sin\phi}{1-\cos\phi}
    \{I_{s}(t-r/c)-I_{s}[t-r/c-(L/2c)(1-\cos\phi)]\}
    \notag
    \\&+\frac{\sin\phi}{1+\cos\phi}
    \{I_{s}(t-r/c)-I_{s}[t-r/c-(L/2c)(1+\cos\phi)]\}),
    \label{E_dipole}
\end{align}
where $\phi$ is the observation angle with respect to the wire axis and $L$ is the length of the wire. Since the radiation energy peaks in the wire-normal direction as can be found in Eq.(\ref{E_dipole}), coupling of the dipole radiation into SPPs on the wire is weak. Thus, the models of current-carrying line antenna based on return currents\cite{JianshuoWang2024RadiationDA} also do not apply to our results.

\textbf{Mearsurement of expelled hot electrons.} The discussions mentioned above demonstrate that the bidirectional THz SPPs are not produced by the guided hot electrons\cite{53.11NP2017,55.nature2022,54.PRE2017,66.28OE2020}, the return currents \cite{JianshuoWang2024RadiationDA}, or the backside hot electrons\cite{10.1063/5.0013415,PhysRevE.108.065211,46.prx2020}. We look at the initial angular distributions of the laser expelled hot electrons there. Fig.~\ref{fig:3}(b) shows that $>80\%$ of the electrons are expelled in the wire normal direction, only a very small number are guided along the wire. Thus, the THz SPPs could already have been generated by the electrons ejected from the LIS in the wire normal direction.

\section{Simulation and Discussion}

\textbf{Simulation Setup.} We use a two-dimensional PIC program from OSIRIS\cite{61.2331sbh2002} to simulate generation of the hot electrons and THz SPPs. A three-dimensional finite-difference time-domain code from CST\cite{62.CST} Particle Studio is also used to simulate generation and delivery of THz SPPs. 
In the PIC simulation, a subrelativistic laser pulse of normalized intensity $a_0=eE/m\omega c=0.3$, wavelength $\lambda_0=800$ nm, FWHM duration 30 fs, and focal spot $6\lambda_0$ normally irradiates the wire at its midpoint from the top. The wire is of diameter $D=32$ $\mu$m (or $40\lambda_0$) and density $n=20n_c$, where $n_c$ is the critical density. It has an inhomogeneous preplasma of width $20\lambda_0$ and scale length $L\sim5\lambda_0$. The PIC results on the laser expelled hot electrons are used as input in the CST simulation, namely, the transversely emitted electrons from the 30 $\mu$m-diameter main wire is of charge $0.5n_c$, pulse duration 2.4 ps, velocity $0.5c$ (or energy 80 keV), and divergence angle $30^\circ$.

\textbf{Agreements between Simulation and Experiments.} Our simulation results are in good agreement with that of the experiments. We found that about $88\%$ of the total number of hot electrons are expelled in the wire normal direction and the rest propagate along the wire in both directions, it agrees with the experiment results shown in Fig.~\ref{fig:3}(b). The spectra of the THz SPPs from the PIC and CST simulations shown in Fig.~\ref{fig:1}(c) also match our experiment results after convoluting the temporal resolution ($\sim800$fs) of the electro-optics measurements\cite{63.73apl1998}. 

\begin{figure}[ht]
\centering
\includegraphics[width = 0.9\textwidth]{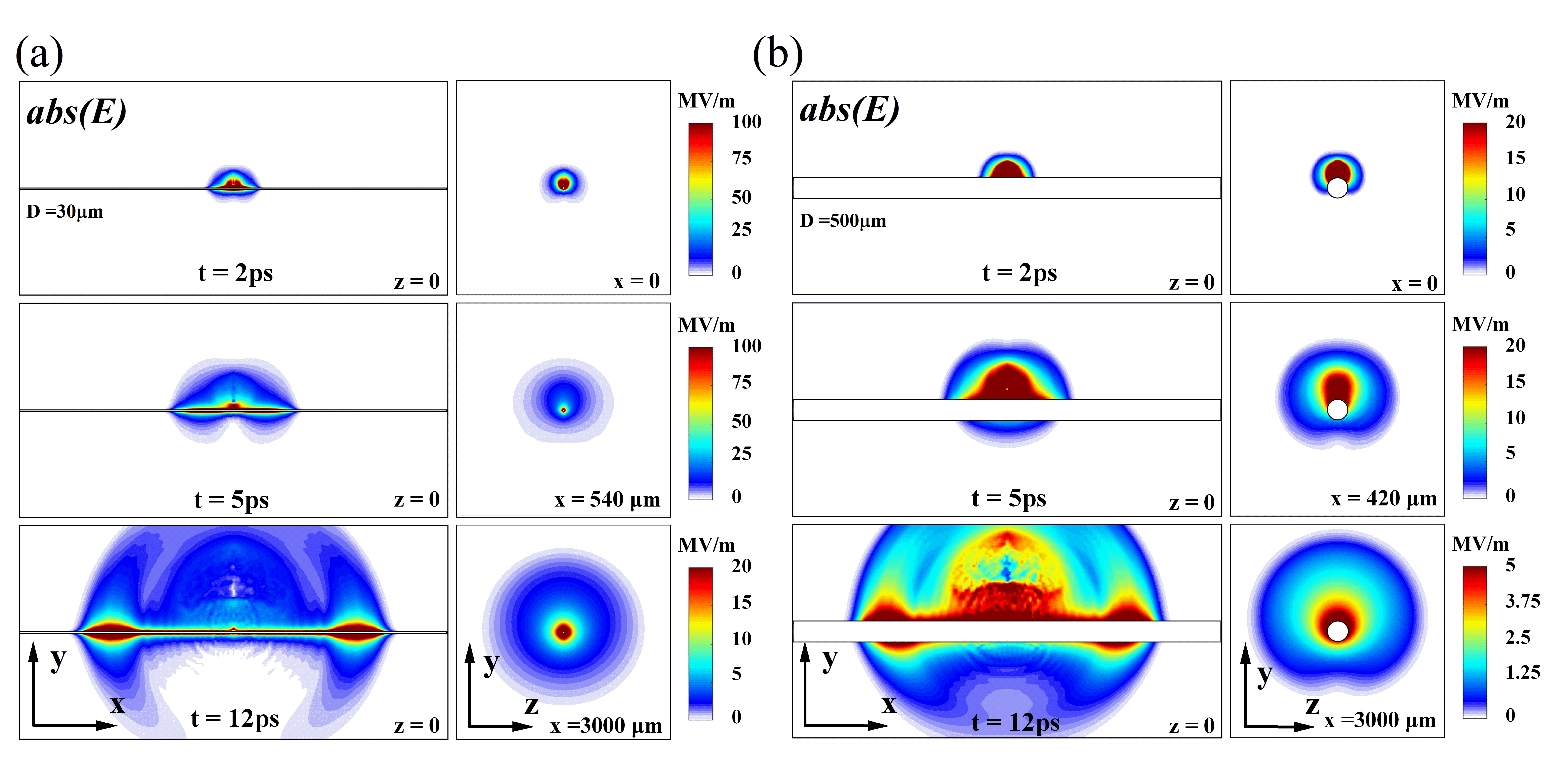}
\caption{\label{fig:4}CST simulation results. Evolution of the radial electric field $|E|$ ($abs(E)$) of the THz SPPs excited by the hot electron bunches along (in the $x$-$y$ plane) and transverse to (in the $z$-$y$ plane) the wire axis, for wire diameters (a) $D=30$ $\mu$m and (b) $500$ $\mu$m.}
\end{figure}

\textbf{Excitation of SPPs.} Fig.~\ref{fig:4} shows that a highly-localized electric fields is generated outside the wire by the laser-expelled electrons. (The transverse field profile shows that there is no field deep inside the thin wire centered at $r=0$.) Radially polarized SPPs are created and propagated in both directions along the wire. The evolution is delayed if a metal wire of larger diameter is used, see Supplementary Material. This scenario suggests that the SPPs are not directly excited by the laser-expelled hot electrons but are transformed from the free-space THz radiation produced by CTR. 
The CTR-produced free-space THz radiation is radially polarized and coupled into SPPs propagating along the wire, which is a Sommerfeld surface mode that is free from coupling loss in the transformation, as occurring in other scenarios \cite{48.JOSAB2005,20.oe2006,50.NC2016,51.PRB2008}.

\begin{figure}[ht]
\centering
\includegraphics[width = 0.9\textwidth]{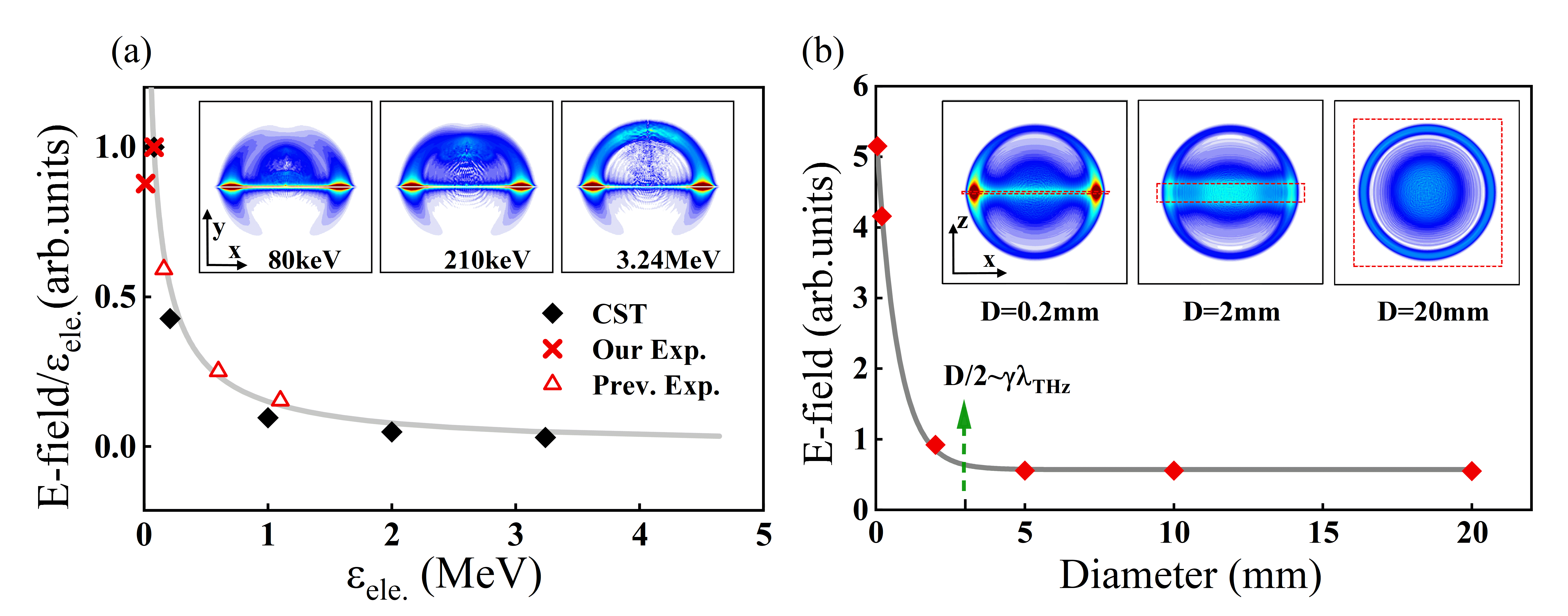}
\caption{\label{fig:5}(a) Dependence of the maximum electric field of the THz SPPs obtained in simulation (CST, at 13 ps after fast-electron ejection) and experiments (Exp.)\cite{55.nature2022,53.11NP2017,66.28OE2020,JianshuoWang2024RadiationDA} (normalized by electron kinetic energy $\epsilon_e$) on the electron kinetic energy, fits well the analytical CTR formula integrated over the $10^\circ$ segment centered at the wire axis (Eq.(\ref{CTR of an electron beam}))\cite{64.93pre2016,65.10pop2003}. The diameter of the wire is 50 $\mu$m. The inset shows the distribution in the $x$-$y$ plane of the electric field magnitude of the THz radiation for electron kinetic energy $\epsilon_e=80$ keV, 210 keV, and 3.24 MeV. (b) Dependence of the maximum electric field of the SPPs on the wire diameter $D$ at 25 ps after hot-electron ejection, which is well fitted by the function $1+Ae^{-a(D/2)/\gamma\lambda}$, where the second term is from slit diffraction radiation and $\gamma\lambda\sim1.7$ mm \cite{70.107JETP2008,71.bookDR}. The wire cross-section areas are 50 $\mu$m$\times$(0.2, 2, and 20) mm. The insert shows the distribution in the $x$-$z$ plane of the electric field magnitude of the THz radiation. The wire boundaries are marked by red dashed lines.}
\end{figure}

\textbf{Influence of electron kinetic energy.} The transformation of CTR-induced THz radiation into SPPs free from coupling loss requires excitation by subrelativistic electrons. 
As illustrated in Fig.~\ref{fig:5}(a), the normalized electric field of the THz SPPs, obtained in our simulation
and experiments, as well as previous experiments \cite{53.11NP2017,66.28OE2020,JianshuoWang2024RadiationDA}, attenuate sharply when the electron kinetic energy increases from subrelativistic (80 keV) to relativistic (3 MeV). 
However, under the laser intensity droping to $1.5\times 10^{16} W/cm^2$, much lower laser absorption results the THz energy blow the measurement threshold.
A considerable part of the THz radiation is emitted into the free space perpendicular to the wire, which is consistent with the CTR theory that the radiation approaches the ejection direction of relativistically energetic electrons\cite{64.93pre2016,39.116prl2016,65.10pop2003,34.pre2004}, resulting in poor coupling with the wire guided SPPs.
Since for wire length much greater than its diameter, the CTR energy spectrum in the axial direction can be expressed as (similar to that of CTR by electron beam crossing the surface of an infinite plane)\cite{10.10631.1491413}
\begin{equation}
    \frac{d^2\varepsilon}{d\omega d\Omega}=
    \frac{N(N-1)e^2}{\pi^2c}
    (\frac{\beta\sin\phi}{1-\beta^2\cos^2\phi})^2\times e^{-(\frac{2\pi}{\lambda})^2\times[(a\sin\phi)^2+(\tau_{e}c)^2]},
    \label{CTR of an electron beam}
\end{equation}
where the bunch of $N$ electron is of Gaussian profile in both the transverse and longitudinal directions, $a$ and $\tau_{e}$ are its radius and pulse duration, respectively, $\beta$ is the electron velocity normalized by the light speed, and $\phi$ is the observation angle.
The electric field of SPPs is obtained by integrating Eq. (\ref{CTR of an electron beam}) over $\phi=0$-$10^\circ$ (the CTR over the angle of $10^\circ$ deviates from radially polarized, impeding efficient coupling into SPPs):
\begin{equation}
|\bm{E}_{TR}|^2\sim\int_\phi{\frac{d^2\varepsilon}{d\omega d\Omega}},
\end{equation}
which fits well with the simulation and experimental results and shows that the THz SPPs for relativistic laser incidence\cite{66.28OE2020} is of the lower conversion efficiency compared with that for subrelativistic laser incidence\cite{55.nature2022,53.11NP2017} .

\textbf{Effect of wire radius.} We have also considered the effect of the wire radius $D/2$ on the THz SPPs. As shown in Fig.~\ref{fig:5}(b), the THz SPPs increases dramatically at $D/2\sim\gamma\lambda\sim1.7$ mm, since in our case the Lorentz factor is $\gamma=1.15$ (for electron of speed $0.5c$), and the wavelength of the THz radiation is $\lambda=1.5$ mm. This result is in good agreement with that of the CTR model\cite{67.169NIMB1998,68.227NIMB2005,69.90JETP2000}.
When the radius of the wire is smaller than $\gamma\lambda$, i.e., comparable with that of the transverse extent of electrons' self-fields, the diffraction radiation from the wire edge can suppress the transverse CTR of the SPPs, and enhance the axial CTR. The latter originates from modulation of the diffraction radiation, so that amplitude variation of the SPPs along the wire follows that of slit diffraction radiation $e^{-a(D/2)/\gamma\lambda}$\cite{70.107JETP2008,71.bookDR}, see Supplementary Material.

\textbf{Explaining the Previous Experiments with CTR.} Amplification of THz SPPs along wire reported in previous studies\cite{55.nature2022,53.11NP2017} can also be explained in terms of the CTR scenario here. There, diffraction radiation can also suppress the CTR along the wire. When the wire length varies from subwavelength to $\gamma\lambda$, diffraction radiation from the wire end decreases rapidly, and the CTR-generated THz radiation increases until the wire length reaches $\sim\gamma\lambda\sim1.95$ mm\cite{53.11NP2017} following the rule of $-e^{(-D/2)/\gamma\lambda}$. 
The near-field result on THz SPPs\cite{55.nature2022} is also consistent with that of the CTR: the radiation formation length, or coherence length\cite{72.25prl1970,39.116prl2016} is $l_f\sim\beta\lambda=0.6$ mm since $\beta\sim0.63$ and $\lambda=0.94$ mm, which corresponds to the 2 ps THz SPPs in their experiment.

\section{Summary}
We have considered generation and propagation of THz SPPs and wire guided hot electrons on a long subwavelength-radius metal wire irradiated by subrelativistic femtosecond laser pulses. Correlation analyses of the THz SPPs and hot electrons show that intense THz SPPs can be efficiently produced from the CTR excited by the hot electrons ejected from the wire by the subrelativistic laser pulse. That is, they are not from the guided fast electrons or the return current propagating along the wire considered previously, nor from the relativistic electrons directly accelerated by a relativistic laser.
The present intense THz SPPs generation scheme has the advantage of only readily-available subrelativistic laser is needed, structural simplicity, low transmission loss, negligible group-velocity dispersion of the SPPs, as well as bidirectional delivery, so that it should be useful for integrated THz devices such as endoscopic THz systems with subwavelength spatial resolution, nonlinear THz pump-probe detection, waveguide-based multistage THz accelerators, etc.\cite{73.6NCl2015,74.12NP2018,75.120prl2018,76.352science2016,77.8prx2018}.

\begin{acknowledgments}
We thank L. Q. Yi and B. F. Shen for their helpful discussions. This work was supported by the National Natural Science Foundation of China (Grant Nos. 12175230, 12225505, 11775223, 12205298, and 12235014), the Strategic Priority Research Program of the Chinese Academy of Sciences, China (Grant No. XDB16), and Fundamental Research Funds for the Central Universities of China.
\end{acknowledgments}

%

\end{document}